\pdfoutput=1
\documentclass[aps,pra,showpacs,reprint,groupedaddress]{revtex4-1}
\usepackage{graphicx}
\usepackage{amsmath}
\usepackage{amssymb}

\begin{document}
\title[Structural physical approximations and entanglement
witnesses]{Structural physical approximations and entanglement witnesses
}
\author{Bang-Hai Wang$^{1,2}$}%
\author{Dong-Yang Long$^2$}%

\affiliation{%
$^1$School of Computers, Guangdong University of Technology, Guangzhou 510006, People's Republic of China,\\
$^2$Department of Computer Science, Sun Yat-sen University, Guangzhou 510006, People's Republic of China
}%
\date{\today}

\begin{abstract}
The structural physical approximation (SPA) to a positive map is considered to be one of the most important methods to detect entanglement in the real physical world. We first show that an arbitrary entanglement
witness (EW) $W$ can be constructed from a separable density
matrix $\sigma$ in the form of $W=\sigma-c_{\sigma} I$, where $c_{\sigma}$ is a non-negative number and $I$ is the identity
matrix.
Following the general form of EWs from separable states, we show a sufficient condition and a sufficient and necessary condition in low dimensions of that SPAs to positive maps do not define entanglement-breaking channels. We show that either the SPA of an EW or the SPA of the partial transposition of the EW in low dimensions is an entanglement-breaking channel. We give sufficient conditions of violating the SPA conjecture [\emph{Phys. Rev. A }{\bf 78,} 062105 (2008)]. Our results indicate that the SPA conjecture is independent of whether or not positive maps are optimal.
\end{abstract}
\pacs{03.65.Ud, 03.67.-a }

\maketitle

\section{Introduction}

Quantum entanglement is a central notion in
the field of quantum information \cite{Horodecki09,Guhne09}. It allows us to
realize various types of quantum information processing, which are not
achievable in classical physics, such as quantum
computation \cite{Shor94}, quantum dense coding \cite{Bennett92},
quantum teleportation \cite{Bennett93}, quantum
cryptography \cite{Ekert91}, etc. However, quantum entanglement is still not fully understood by researchers. There exists no effective method to detect a given state is entangled or not.

To the best of our knowledge, the most powerful method to detect entanglement up to date may be the one based
on the notion of positive maps \cite{M.Horodecki96,Augusiak11}, i.e.,
a given state $\rho$ on Hilbert space $\mathcal{H}_A\otimes\mathcal{H}_B$ is separable if and only if
$(I\times\Lambda)\rho$ is positive for an arbitrary positive but not completely
positive (PNCP) map $\Lambda$ : $\mathcal{H}_B\rightarrow \mathcal{H}_A$. Then obviously it is critical to find ways to achieve experimental detection of entanglement with a PNCP
map, which
does not represent physical process. The earliest \cite{Augusiak11} and most important work may be
the structural physical approximation (SPA) \cite{Horodecki03,Horodecki02}. It is based on the idea
that a positive map will result in a completely positive (CP) map
when it is mixed with a simple CP map. A CP map presents
physical process and can be implemented experimentally. In addition, the resulting map
keeps the structure of the output of the nonphysical map $\Lambda$.

For our purpose, we can only consider the quantum states on the finite
dimensional Hilbert space
$\mathcal{H}_{AB}=\mathcal{H}_A\otimes\mathcal{H}_B$. We let
dim$(\mathcal{H}_{A})={d}_{A}$, dim$(\mathcal{H}_{B})={d}_{B}$ and
dim$(\mathcal{H}_{AB})={d}_{AB}$. Here we study whether or not the structural approximation to a positive map $\Lambda$ : $\mathcal{H}_A\rightarrow \mathcal{H}_B$ defines an entanglement-breaking (EB) channel. The structural approximation is obtained through minimal admixing of white noise
\begin{equation}
\tilde{\Lambda}(\rho)=p\text{tr}(\rho)\frac{I}{d_B}+(1-p)\Lambda(\rho).
\end{equation}
In other words for \emph{minimal admixing}, we take the smallest noise probability $0<p<1$ for which $\tilde{\Lambda}$ become completely positive. As a consequence of the Jamio{\l}kowski-Choi isomorphism \cite{Jam}, we are led to study the separability of entanglement witnesses (EWs) of the form
\begin{equation}
\tilde{W}_\Lambda=I\otimes\tilde{\Lambda}(P_+)=\frac{p}{d_{AB}}I+(1-p)W_\Lambda\label{SPAstatetoPM}
\end{equation}
for minimal $p$ such that $\tilde{W}_\Lambda\geq0$, where $P_+=|\beta\rangle\langle\beta|$, $|\beta\rangle=d_{A}^{-1/2}\sum_{i}|i\rangle\otimes|i\rangle$ on
$\mathcal{H}_A\otimes\mathcal{H}_A$. For simplicity, in the following we shall freely use the positive map and the EW.

It was first noticed by Fiur$\acute{a}\breve{s}$ek that the SPA of the (optimal) partial transposition map (PT) map $I\otimes T$ in the two-qubit case is an EB channel \cite{Fiurasek02}. Recently, it was observed in many examples of optimal positive maps that their SPAs are EB by Korbicz \emph{et al.} They formulated a conjecture (called the SPA conjecture) that SPAs to optimal positive maps are EB \cite{Korbicz08}. Very recently, some examples constructed show that the SPA conjecture fails \cite{Ha12a,Stormer12}.

Even though the fact that the SPA conjecture fails for certain optimal EWs has been shown, some problems related to the conjecture remain unknown. In this paper, we first show that an arbitrary EW $W$ can be written in the form of $W=\sigma-c_{\sigma} I$, where
$\sigma$ is a separable density matrix, $c_{\sigma}$ is a non-negative number, and $I$ is the identity
matrix. Following the general form of EWs from separable states, we give a simple form of the SPA of EWs. Based on the positive partial transposition (PPT) separability criterion \cite{Peres96}, we show a sufficient condition and a sufficient and necessary separability condition in low dimensions ($d_A\times d_B\leq6$) of that structural approximations to positive maps do not define entanglement-breaking channels. We find that either the SPA of an EW or the SPA of the partial transposition of the EW in low dimensions is an EB channel. We show sufficient conditions of violating the SPA conjecture. We show that the SPA conjecture does not need to be based on the optimality of positive maps.

\section{Constructing All Entanglement Witnesses from Separable Density Matrices}

EWs are observables
that completely characterize separable (not entangled) states and
allow us to detect entanglement physically \cite{Horodecki09}. An observable $W=W^\dag$ is called an EW if (i) $\text{tr}(W\sigma)\geq0$ for an arbitrary separable state $\sigma$; and (ii) there exists an entangled state $\pi$ to make $\text{tr}(W\pi)<0$. To balance out possible unnormalization of quantum states, another property of
EWs is required: (iii) if $W$ is an EW, $\gamma W$ keeps all
properties of $W$ as an EW for a non-negative
number $\gamma$. In this case, we say that $\gamma W$ is the same EW as $W$ ( or $\gamma W$ is as fine as $W$ \cite{Sperling09}). To compare the action of
different EWs, Lewenstein \emph{et al.} defined (iii') $\text{tr}(W)=1$ in \cite{Lewenstein00}. Note that we will encounter interchangeably incompatible (iii) and (iii') on different cases.

Following the definition in Ref. \cite{Lewenstein00}, we have: (i)
Given an EW $W$, $D_W=\{\pi\geq0$, such that $\text{tr}(W\pi)<0\}$, i.e.,
the set of density matrices detected by $W$; (ii) Given two EWs,
$W_1$ and $W_2$, $W_2$ is finer than $W_1$ if $D_{W_1}\subseteq
D_{W_2}$; (iii) $W$ is an optimal entanglement witness (OEW) if
there exists no other finer EW. A necessary condition for an EW $W$ to be optimal is that there must be a separable $\sigma$ with $\text{tr}(W\sigma)=0$ \cite{Guhne09}. It is, however, not a sufficient condition \cite{Terhal02,Brandao05,Thanks}. Following the definition in Ref. \cite{Badziag07}, we have (iv) $W$ is a weakly optimal entanglement witness (WOEW) if its expectation value vanishes on at least one product vector. Clearly, an OEW is a special WOEW.

\subsection{Constructing all entanglement witnesses from separable density matrices}

As a matter of fact, EWs provide one of the main methods and the best known one for
entanglement detection in composite quantum system \cite{Horodecki09,Sperling09,Doherty04,Hou10a,Chruscinski09,Chruscinski10}. Although common entanglement witnesses for some entangled states have been studied \cite{Wu07,Ganguly13}, given an entangled state, it is difficult to construct an EW. Unfortunately, it is a nondeterministic polynomial-time
(NP) hard problem to determine EWs for all entangled states \cite{Doherty04,Hou10a,Gurvits04}.

Our recent work \cite{Wang11} showed that an arbitrary EW $W$ can be written as
\begin{equation}
W=\rho-c_\rho I\,\label{ew-form0}
\end{equation}
where $\rho$ is a (unnormalized) density matrix and $c_\rho>0$
is a real number related to $\rho$.
For example, $\rho=\frac{p}{d_{AB}}I+\frac{1-p}{d_A}\sum_{ij}|i\rangle|j\rangle\langle
j|\langle i|$ is a density matrix for $\frac{d_{AB}}{d_A+d_{AB}}\leq p\leq1$, where the range of $p$ is determined by the positivity of the state, and
\begin{equation}
W=\rho-\frac{p}{d_{AB}}I=\frac{1-p}{d_A}V (\frac{d_{AB}}{d_A+d_{AB}}\leq p<1)
\end{equation} is the same EW as $V$, where $V$ is the swap operator \cite{Werner89}. The form in Eq. (\ref{ew-form0}) of some EWs were investigated in Ref. \cite{Acin01,Sanpera01}. It was shown that the EW of a given entangled state $\pi$ can be written as
\begin{eqnarray}
W=\sigma-\pi-\text{tr}[\sigma(\sigma-\pi)]I,\label{EW_Pitt}
\end{eqnarray}
where $\sigma$ is the nearest separable state to $\pi$ \cite{Pittenger02,Pittenger03}. However, such a state $\sigma$ is difficult to compute.

By Eq. (\ref{ew-form0}) and the trace inequality, we have showed an EW for a given state could be constructed from a commuting state \cite{Wang11}. According to the property of EWs and the relation between an arbitrary
density matrix and a separable density matrix, we find that the
general form of EWs in Eq. (\ref{ew-form0}) can be written as the form
related to separable density matrices.

\textbf{Theorem 1.} An arbitrary bipartite density matrix $\pi$ is entangled
if and only if there exists a separable state $\sigma$ and a
non-negative number $c_\sigma$ such that the operator
\begin{equation}
W=\sigma-c_\sigma I\label{ew-density0}
\end{equation}
satisfies $\text{tr}(W\pi)<0$ and $\text{tr}(W\sigma')\geqslant 0$ for all
separable states $\sigma'$.

 \textbf{Proof:} Following Eq. (\ref{ew-form0}) for an arbitrary EW $W'$, there exists a density
 matrix $\rho$ such that
\begin{equation}
W'=\rho-c_\rho I\label{ew-density1}.
\end{equation}

Observe the depolarizing channel \cite{Nielsen00}
\begin{equation}
\tilde{\rho}_p = (1-p)\frac{I}{d_{AB}} +p\rho (0<p<1)\label{depolarixingchannel}.
\end{equation}

By Eq. (\ref{depolarixingchannel}),
\begin{align}
p\rho &=\tilde{\rho}_p-(1-p)\frac{I}{d_{AB}}\\
p\rho-pc_\rho I &=\tilde{\rho}_p-(1-p)\frac{I}{d_{AB}}-pc_\rho I\\
p(\rho-c_\rho I)&=\tilde{\rho}_p-[pc_\rho+\frac{1-p}{d_{AB}}]I.
\end{align}

By Eq. (\ref{ew-density1}),
\begin{align}
pW'&=\tilde{\rho}_p-[pc_\rho+\frac{1-p}{d_{AB}}]I\label{SameEW}.
\end{align}

It is known that the right-hand side of equality in Eq.
(\ref{SameEW}) is the same EW as $W'$ for $0<p<1$ \cite{Wang11}. Considering the
right-hand side of equality in Eq. (\ref{SameEW}), for
sufficiently small values of $p\leq p_s$, the state $\tilde{\rho}_p$
becomes a separable state $\sigma$ for an arbitrary
$\rho$ \cite{Braunstein99,Zyczkowski98,Vidal99,Gurvits02}. Therefore,
for an arbitrary EW $W'=\rho-c_\rho I$, there exists the same EW as $W'$,
$W=\sigma-c_\sigma I$, where $\sigma$ is a separable state and
$c_\sigma$ is a non-negative real number.\hfill$\blacksquare$

It is a simple result, since the identity is of full rank. However, there are some interesting consequences can be derived from it.

\textbf{Corollary 1.} If $W=\sigma-c_\sigma I$ is an EW,
\begin{equation}
 \lambda_{0\sigma}<c_\sigma,
\end{equation}
where $\lambda_{0\sigma}$ denotes the minimum
eigenvalue of $\sigma$.

 \textbf{Proof:} If $\sigma=\sum_{i=0}^{k}\lambda_{i\sigma}P_i$ is the spectral decomposition of $\sigma$, $W=\sigma-c_\sigma I=\sum_{i=0}^{k}(\lambda_{i\sigma}-c_\sigma)P_i$ is the spectral decomposition of $W$, where $\sum_{i=0}^{k}P_i=I$. If $\lambda_{0\sigma}\geq c_\sigma$, $W=\sigma-c_\sigma I=\sum_{i=0}^{k}(\lambda_{i\sigma}-c_\sigma)P_i>0$, which is impossible, since $W<0$. \hfill$\blacksquare$

\subsection{A general method to construct finer and weakly optimal entanglement
witnesses}

\textbf{Corollary 2.} An EW $W=\sigma-c_{\sigma} I$ is weakly optimal if and only if $c_{\sigma}=c_\sigma^{max}$, where
\begin{equation}
 c_\sigma^{max}=\inf_{\parallel\mu_A\parallel=1,
\parallel\nu_B\parallel=1}\langle\mu_A\nu_B|\sigma|\mu_A\nu_B\rangle,
\end{equation}
is the maximum number in
$c_\sigma$ which makes $W=\sigma-c_\sigma I$ an EW and $|\mu_A\nu_B\rangle$ is an arbitrary unit product vector \cite{Wang11}.

By Corollary 2, we can give a method to construct a weakly optimal entanglement witnesses for $W=\sigma-c_\sigma I$ by computing $c_\sigma^{max}$. However, it is not easy to compute $c_\sigma^{max} $\cite{Wang11,Leinaas10}. Since an arbitrary EW can be written in the form of $W=\sigma-c_\sigma I$, we obtain the following result by Corollary 2.

\textbf{Corollary 3.} For an arbitrary OEW $W^{opt}$, there exists a separable density matrix $\sigma$ such that $W^{opt}=\sigma-c_\sigma^{max}I$.

In other words, an arbitrary OEW $W$ can be written in the form of $W=\sigma-c_\sigma^{max}I$, but the form of an EW $W=\sigma-c_\sigma^{max}I$ may be not optimal.

\textbf{Corollary 4.} If a non-weakly-optimal EW $W=\sigma-c_{\sigma} I$ can detect
entangled state $\pi$, $W_i=\sigma-c_{i\sigma} I$ can also detect $\pi$,
where $c_{\sigma}<c_{i\sigma}\leq c_\sigma^{max}$, $\{i=1,2,\cdots\}$
refers to the index set.

 \textbf{Proof:}
 (i) Since $W=\sigma-c_{\sigma} I$ can detect the
entangled state $\pi$, we have
\begin{align}
\text{tr}(W\pi)&=\text{tr}(\sigma\pi)-c_{\sigma}<0,\\
\text{tr}(\sigma\pi)&<c_{\sigma}\label{traceEW}.
\end{align}
By Eq. (\ref{traceEW}) and $c_{\sigma}<c_{i\sigma}$, we have
\begin{align}
\text{tr}(\sigma\pi)&<c_{i\sigma},\\
\text{tr}(W_i\pi)&=\text{tr}(\sigma\pi)-c_{i\sigma}<0.
\end{align}

(ii) Since $W^{wopt}=\sigma-c_\sigma^{max} I$ is a weakly optimal EW by Corollary 2,
\begin{equation}
 \text{tr}(W^{wopt}\sigma')\geq0
\end{equation}
 and
\begin{equation}
 \text{tr}(\sigma\sigma')\geq c_\sigma^{max}
\end{equation}
for all separable states $\sigma'$.

By $c_{i\sigma}\leq c_\sigma^{max}$,
\begin{equation}
\text{tr}(W_i\sigma')=\text{tr}(\sigma\sigma')-c_{i\sigma}\geq0,
\end{equation}for all separable states $\sigma'$.

Therefore, $W_i=\sigma-c_{i\sigma} I$ is an EW
and it can detect $\pi$ for $c_{\sigma}<c_{i\sigma}\leq c_\sigma^{max}$.\hfill$\blacksquare$

By Corollary 4, we can construct a finer EW than $W=\sigma-c_{\sigma} I (\lambda_{0\sigma}<c_{\sigma}<c_\sigma^{max})$ through adding an minor number to $c_{\sigma}$.

 \textbf{Corollary 5.} A non-weakly-optimal EW $W_2=\sigma-c_{2\sigma} I$ is finer than $W_1=\sigma-c_{1\sigma} I$ if
 there exists $0<\delta\leq c_\sigma^{max}-c_{2\sigma}$ such that $c_{1\sigma}+\delta=c_{2\sigma}\leq c_\sigma^{max}$.

By Corollary 5, we can immediately give a method to construct finer EWs for a given non-weakly-optimal EW. Note that it is not so easy to construct finer EWs for a weakly-optimal-but-not-optimal EW. General method of optimization of EWs can be found in Ref. \cite{Lewenstein00}.

A violation of a Bell inequality for a bipartite density matrix $\pi$, considered within quantum mechanics, can be reformulated as an EW $W_\pi$ for entangled $\pi$ \cite{Terhal00}. Based on Corollaries 3, 4 and 5, it is not difficult to understand relations between such Bell inequalities and EWs in Ref. \cite{Hyllus05}. The relation was revealed that how much an OEW has to be shifted by adding the identity operator to make it positive on all states admitting a local hidden variable model and how much can be subtracted from a CHSH witness \cite{Clauser69}, while preserving the EW properties.
Theorem 1 shows that
the research on separable density matrices can replace the research on
entanglement witnesses. Generally, it is not easy to characterize the separable state $\sigma$ with $W=\sigma-c_\sigma I$ being an EW.




We illustrate our results for the case of the qubit state. Observe the EW
\begin{eqnarray}
W&=&
    \left[
      \begin{array}{cccc}

                 0.1 &0       & 0      &0  \\

                 0   &0.4     &0.1+0.4 & 0 \\

                 0   &0.1+0.4 &0.4     &0 \\

                 0   &0       &0       &0.4
      \end{array}
    \right]\label{SpecialCase}\\
    &=&\rho-0.1I\\
    &=&\sigma-0.4I,
\end{eqnarray}where
\begin{equation}
\rho=\left[\begin{array}{cccc} 0.2 & 0 & 0 & 0\\
                        0 & 0.5 & 0.5 & 0\\
                        0 & 0.5 & 0.5 & 0\\
                        0 & 0 & 0 & 0.2 \end{array}\right], \quad
\sigma=\left[\begin{array}{cccc} 0.5 & 0 & 0 & 0\\
                            0 & 0.8 & 0.5 & 0\\
                            0 & 0.5 & 0.8 & 0\\
                            0 & 0 & 0 & 0.5 \end{array}\right],
\end{equation}
$\rho$ is (unnormalized) entangled, $\sigma$ is (unnormalized) separable. We can compute $\lambda_{0\rho}=0, \lambda_{0\sigma}=0.3$. We can compute $c_\rho^{max}=0.1, c_\sigma^{max}=0.4$ by the method in Appendix A. Thus, $W_1=\rho-0.05I$ and $W_2=\sigma-0.35I$ are EWs. $W_1$ is the same EW as $W_2$. $W$ are finer than $W_1$ and $W_2$. We will see below that Eq. (\ref{SpecialCase}) is a special case of Eq. (\ref{exampleKorbicz}) for $a=0.1, b=0.4$.

\section{Structural approximations to positive maps and entanglement-breaking channels}

An equivalent presentation of the SPA conjecture
is that SPAs to optimal entanglement witnesses correspond to
separable (unnormalized) states, i.e., the SPA to an optimal EW
$W^{opt}$,
\begin{eqnarray}
\tilde{W}^{opt}=\frac{p}{d_{AB}}I+(1-p)W^{opt} (0<p<1)
\end{eqnarray}
is separable where $p$ is the smallest parameter for which $\tilde{W}^{opt}$
is a positive operator (possibly unnormalized state) \cite{Chruscinski09}.

Let $s(p)=\frac{p}{(1-p)d_{AB}}$.
By considering possible unnormalization of
states, the SPA conjecture can be simplified as follows:
\begin{eqnarray}
\tilde{W}^{opt}=W^{opt}+sI\label{SPAstatetoOpt}
\end{eqnarray} is separable where $s=\frac{p}{(1-p)d_{AB}}>0$ is the smallest parameter for which $\tilde{W}^{opt}$
is a positive operator (possibly unnormalized state) since $s(p)$ is a monotonically increasing function for $p$ in the interval $(0,1)$.

Similarly, the SPA for a general EW $W$ (optimal or not) can be simplified as
\begin{eqnarray}
\tilde{W}=W+sI\label{SPAstatetogeneral}.
\end{eqnarray}

\textbf{Theorem 2.} The SPA to an arbitrary EW $W=\sigma-c_\sigma I$ (optimal or not) corresponds to a separable state if and only if
\begin{equation}
\tilde{W}=\sigma-\lambda_{0\sigma} I
\end{equation}
is separable, where $\lambda_{0\sigma}$ is the minimum eigenvalue of $\sigma$.

\textbf{Proof:} An arbitrary EW $W$ can be written as $W=\sigma-c_\sigma I$ by Theorem 1. By Corollary 3 and Eq. (\ref{SPAstatetoOpt}), we have

\begin{eqnarray}
\tilde{W}^{opt}&=&\sigma-c_\sigma^{max} I +sI\\
               &=&\sigma-(c_\sigma^{max}-s)I.
\end{eqnarray}

For taking the smallest parameter $s$ to make $\tilde{W}$ being positive, $c_\sigma^{max}-s=\lambda_{0\sigma}$, that is

\begin{equation}
\tilde{W}^{opt}=\sigma-\lambda_{0\sigma} I.
\end{equation}

Similarly, we can also obtain the above result by Eq. (\ref{SPAstatetogeneral}).

Therefore, the separability problems of SPA to both optimal maps and non-optimal maps become the same problem, that is, whether the $\tilde{W}=\sigma-\lambda_{0\sigma} I$ is separable or not for either $W=\sigma-c_\sigma^{max} I$ being an WOEW (including OEW) or $W=\sigma-c_\sigma I$ being a non-weakly-optimal EW. \hfill$\blacksquare$

\textbf{Corollary 6.} If $W=\sigma-c_\sigma I$ is an EW with $\sigma$ being not full rank, its SPA defines an entanglement-breaking channel (EBC) (the output is just $\sigma$).

Unless otherwise specified, EWs with $W=\sigma-c_\sigma I$ discussed below refer to EWs with $\sigma$ being full rank.

\textbf{Theorem 3.} If $\lambda_{0W^\Gamma}<\lambda_{0W}$, the SPA of an arbitrary $W$ does not correspond to a separable state, where $\Gamma$ refers to partial transposition.

\textbf{Proof:} Suppose $W=\sigma-c_\sigma I$ by Theorem 1. If $\lambda_{0W^\Gamma}<\lambda_{0W}$, $\lambda_{0\sigma^\Gamma}<\lambda_{0\sigma}$, and $\tilde{W}^\Gamma=\sigma^\Gamma-\lambda_{0\sigma} I<0$. $\tilde{W}$ is not separable by PPT criterion \cite{Peres96}. \hfill$\blacksquare$

\subsection{Structural approximations to entanglement witnesses in low dimensions}

Following the definition in Ref. \cite{Lewenstein00}, if an EW can be written in the form of $W=P+Q^\Gamma$ with $P, Q\geq0$, we say it decomposable, otherwise we say it indecomposable. It is well known that the division of EWs to
decomposable and indecomposable is translated from positive maps via the Jamio{\l}kowski-Choi isomorphism \cite{Jam}.

\textbf{Corollary 7.} If $W=\sigma-c_\sigma^{max} I$ is a low-dimension OEW,
\begin{eqnarray}
\tilde{W}=\sigma-\lambda_{0\sigma} I
\end{eqnarray} is separable without considering
normalization.

\textbf{Proof:} Since $W=\sigma-c_\sigma^{max} I$ is a low-dimension decomposable OEW, $W^\Gamma\geq 0$ by \cite{Lewenstein00,Korbicz08}, i.e. $(\sigma-c_\sigma^{max} I)^\Gamma\geq 0$.
The minimum eigenvalue of $\sigma^\Gamma$, $\lambda_{0\sigma^\Gamma}\geq c_\sigma^{max}$. By $c_\sigma^{max}>\lambda_{0\sigma}$, $\lambda_{0\sigma^\Gamma}> \lambda_{0\sigma}$.
By PPT criterion in low dimensions \cite{M.Horodecki96}, $\tilde{W}^\Gamma=(\sigma-\lambda_{0\sigma} I)^\Gamma=\sigma^\Gamma-\lambda_{0\sigma} I\geq 0$, and $\tilde{W}$ is separable. \hfill$\blacksquare$

This result indicates that all structural physical approximations to optimal positive maps of
low dimensions define EB channels. It is consistent with the result in Ref. \cite{Korbicz08}. Following Theorem 3 and the necessary and sufficient separability criterion for the case of low dimensions, we have a more general result than the one in Corollary 7.

\textbf{Corollary 8.} The SPA of an arbitrary EW in low dimensions $W$ does not correspond to a separable state if and only if $\lambda_{0W^\Gamma}<\lambda_{0W}$.

In other words, the SPA of an arbitrary entanglement witness in low dimensions $W$ corresponds to a separable state if and only if $\lambda_{0W^\Gamma}\geq\lambda_{0W}$.

\textbf{Lemma 1.} If the partial transposition of an EW $W$, $W^\Gamma<0$, $W^\Gamma$ is also an EW.

\textbf{Theorem 4.} Either the SPA of an EW or the SPA of the partial transposition of the EW in low dimensions is an EB channel.

\textbf{Proof:} Suppose an arbitrary EW $W=\sigma -c_\sigma I$ in low dimensions.
(i) If $\lambda_{0W^\Gamma}\geq\lambda_{0W}$, $\tilde{W}=\sigma-\lambda_{0\sigma} I$ is separable by Corollary 8.
(ii) If $\lambda_{0W^\Gamma}<\lambda_{0W}$, $\lambda_{0\sigma^\Gamma}<\lambda_{0\sigma}$, and $W^\Gamma$ is an EW by Lemma 1. The SPA of $W^\Gamma$ is $\sigma^\Gamma-\lambda_{0\sigma^\Gamma}I$ by Theorem 2. By $(\sigma^\Gamma-\lambda_{0\sigma^\Gamma}I)^\Gamma=\sigma-\lambda_{0\sigma^\Gamma}I$. By $\lambda_{0\sigma^\Gamma}<\lambda_{0\sigma}$, $\sigma-\lambda_{0\sigma^\Gamma}I>0$. By PPT criterion, $\widetilde{W^\Gamma}=\sigma^\Gamma-\lambda_{0\sigma^\Gamma}I$ is separable.
\hfill$\blacksquare$

These results deepen the result in the Fiur\'{a}\u{s}ek [\emph{Phys. Rev. A }{\bf 66,} 052315 (2002)],
which states that the SPA of the (optimal) partial transposition map $I\otimes T$ in the two-qubit case is an EB channel.

Let us see the example in Ref. \cite{Korbicz08} as follows.

The EW
\begin{equation}
W=Q_1+Q_2^\Gamma,\label{exampleKorbicz}
\end{equation}where
\begin{equation}
Q_2=\left[\begin{array}{cccc} a & 0 & 0 & a\\
                        0 & 0 & 0 & 0\\
                        0 & 0 & 0 & 0\\
                        a & 0 & 0 & a \end{array}\right], \quad
Q_1=\left[\begin{array}{cccc} 0 & 0 & 0 & 0\\
                            0 & b & b & 0\\
                            0 & b & b & 0\\
                            0 & 0 & 0 & 0 \end{array}\right]
\end{equation}
with real positive $a$ and $b$. We can compute $\text{tr}[W(|\mu_A\rangle\nu_B\rangle\langle\mu_A\langle\nu_B|)]=0$, where $|\mu_A\rangle=\frac{\sqrt2}{2}(|0\rangle-|1\rangle), |\nu_B\rangle=\frac{\sqrt2}{2}(|0\rangle+|1\rangle)$. All $W$ are WOEWs for $a>0$ and $b>0$. The procedure of computing is presented in Appendix A.

Let $q=1-p$. The SPA of $W$,
\begin{eqnarray}
\widetilde W&=&\frac{p}{4}I+(1-p)\big(Q_1+Q_2^{\Gamma}\big ) \\\label{example}
                       &=&\left[\begin{array}{cccc} qa+\frac{p}{4} & 0 & 0 & 0\\
                            0 & qb+\frac{p}{4} & q(b+a) & 0\\
                            0 & q(b+a) &qb+\frac{p}{4} & 0\\
                            0 & 0 & 0 & qa+\frac{p}{4} \end{array}\right]
\end{eqnarray} is positive for
\begin{equation}\label{condition-pa}
p\ge \frac{4a}{4a+1},
\end{equation}
which is the condition for the structural approximation.

The state (\ref{example}) is not PPT, and hence entangled, for
\begin{equation}\label{condition-pb}
p<\frac{4b}{4b+1}.
\end{equation}

The condition (\ref{condition-pa}) and (\ref{condition-pb}) can be simultaneously
satisfied by taking $b>a$, thus giving a structural approximation which is not EB \cite{Korbicz08}.

In fact, the minimum eigenvalue of $W^\Gamma$, $\lambda_{0W^\Gamma}=b$, and the minimum eigenvalue of $W$, $\lambda_{0W}=a$. If $b>a$, $\lambda_{0W^\Gamma}>\lambda_{0W}$. Since $W^\Gamma<0$ in low dimensions, all $W$ are not optimal for all $a>0, b>0$ ( for $b>a$ and for $b\leq a$ ) \cite{Lewenstein00,Korbicz08}. Interestingly, according to our results, the SPA of $W$ does not correspond to a separable state, but the SPA of $W^\Gamma$ corresponds to a separable if $b>a$; the SPA of $W$ corresponds to a separable state but the SPA of $W^\Gamma$ does not correspond to a separable state if $a\geq b$.

Interestingly, we can easily prove the optimal EW of the EW $W=Q_1+Q_2^\Gamma$ is the same (unnormalized) $Q_2^\Gamma$ either for $b>a$ or for $b\leq a$. The procedure of proof is presented in Appendix B.

Clearly, it is not the reason of the non-optimality of $W$, as stated in Ref. \cite{Korbicz08}, that the SPA of $W$ for $b>a$ is not separable. Thus, we can conclude that the SPA of EWs is independent of the optimality of EWs, and that the SPA conjecture does not need to be based on the optimality of EWs. As already stated in Ref. \cite{Lewenstein00}, we can restrict ourselves to the study of the OEW.

 \subsection{Sufficient conditions of violating the SPA conjecture}

\textbf{Corollary 9.} If $\lambda_{0(W^{opt})^\Gamma}<\lambda_{0W^{opt}}$ (i. e. $\lambda_{0\sigma^\Gamma}<\lambda_{0\sigma}$) for an optimal EW $W^{opt}=\sigma-c_\sigma^{max}I$, $W^{opt}$ violates the SPA conjecture.

This is a sufficient condition of violating the SPA conjecture.

\textbf{Lemma 2 \cite{Lewenstein00,Korbicz08}.} $W$ is an optimal nondecomposable EW (ONEW) if and only if $W^\Gamma$ is an ONEW.

\textbf{Corollary 10.} If $W=\sigma-c_\sigma^{max}I$ is an ONEW with $\lambda_{0W^\Gamma}\neq \lambda_{0W}$, the SPA of $W$ or the SPA of $W^\Gamma$ violates the SPA conjecture.

\textbf{Proof:} Following Theorem 3 and Lemma 2, we can obtain the result. \hfill$\blacksquare$

This is a sufficient condition of violating the SPA conjecture for the ONEW. It is easy to verify. Consider the ONEW in Ref. \cite{Ha12a}
\begin{equation}
W[a,b,c;\theta]= \left[ \begin{array}{ccccccccccc} a     &\cdot   &\cdot  &\cdot  &-e^{i\theta}     &\cdot
&\cdot   &\cdot  &-e^{-i\theta}     \\ \cdot   &c &\cdot    &\cdot    &\cdot   &\cdot &\cdot &\cdot     &\cdot   \\ \cdot  &\cdot    &b &\cdot &\cdot  &\cdot
&\cdot    &\cdot &\cdot  \\ \cdot  &\cdot    &\cdot &b &\cdot  &\cdot    &\cdot    &\cdot &\cdot  \\ -e^{-i\theta}     &\cdot   &\cdot  &\cdot  &a     &\cdot
&\cdot   &\cdot  &-e^{i\theta}     \\ \cdot   &\cdot &\cdot    &\cdot    &\cdot   &c &\cdot &\cdot    &\cdot   \\ \cdot   &\cdot &\cdot    &\cdot    &\cdot
&\cdot &c &\cdot    &\cdot   \\ \cdot  &\cdot    &\cdot &\cdot &\cdot  &\cdot    &\cdot    &b &\cdot  \\ -e^{i\theta}     &\cdot   &\cdot  &\cdot  &-e^{-i\theta}
&\cdot   &\cdot &\cdot  &a \end{array} \right],
\end{equation}
where $a, b, c$ are non-negative real numbers, $\cdot$ denotes $0$ and $-\pi\leq\theta\leq\pi$.

Let $\theta=\pi/12$, $a=\frac{4}{3}cos\frac{\pi}{12}$, $b=\frac{2}{3}cos\frac{\pi}{12}$, and $c=0$. We compute
$\lambda_{0W^\Gamma}\approx-0.6440, \lambda_{0W}\approx-0.7286$, and $\lambda_{0W^\Gamma}\neq\lambda_{0W}$. By Corollary 10, $W[\frac{4}{3}cos\frac{\pi}{12},\frac{2}{3}cos\frac{\pi}{12},0;\pi/12]$ violates
the SPA conjecture.

Since the result of Theorem 3 is followed with the PPT criterion and the PPT criterion is a necessary but not a sufficient
separable condition for separability in higher dimensions, it is not easy to find the necessary condition by our result for that SPA of an arbitrary $W$ does not correspond to a separable state. It is still open whether the SPA of an optimal decomposable entanglement witnesses in higher dimensions is separable or not \cite{Augusiak13}. It indicates that, as separability criteria, there may exist no effective method to detect if the SPA of a given EW is separable or not.

\section{Conclusions}

In summary, we give a general form of an arbitrary EW $W=\sigma-c_{\sigma} I$ from a separable density
matrix $\sigma$.
We show a sufficient condition for that all structural approximations to positive maps define EB channels. For low dimensions as PPT separability criterion, we completely reveal the relation between the SPA to positive maps and EB channels. Our results deepen the works in Fiur\'{a}\u{s}ek [\emph{Phys. Rev. A }{\bf 66,} 052315 (2002)] and in [\emph{Phys. Rev. A }{\bf 78,} 062105 (2008)].
\begin{acknowledgments}
We would like to thank Professor Guang Ping He, Professor Simone Severini, Rui-Gang Du, Ning-Yuan Yao, Dan Wu, Zhang cai, Zhi-Wei Sun, and Hai-Ru Xu for helpful discussions and suggestions. We thank the referee for valuable comments and suggestions to improve the original manuscript. This work is supported by the National Natural Science Foundation of China under Grants No. 61272013 and the National Natural Science Foundation of Guangdong province of China under Grants No. s2012040007302.
\end{acknowledgments}

\section*{Appendix A: The procedure of computing the weakly-optimality of an EW in the two-qubit case}

An arbitrary qubit pure state $|\psi\rangle$ can be written as
$|\psi\rangle=\alpha|0\rangle+\beta|1\rangle$, where $\alpha$ and
$\beta$ are complex number and $|\alpha|^2+|\beta|^2=1$. Because
$|\alpha|^2+|\beta|^2=1$, $|\psi\rangle$ can be rewritten as
\begin{eqnarray}
|\psi\rangle=e^{ir}(\cos{\frac{\theta}{2}}|0\rangle+e^{it}\sin{\frac{\theta}{2}}|1\rangle),
\end{eqnarray}
where $\theta$, $r$ and $t$ are real numbers. The factor of $e^{ir}$
out the front can be ignored since it has no observable
effects \cite{Nielsen00}, and for that reason, $|\psi\rangle$ can be
effectively written as
\begin{eqnarray}
|\psi\rangle=\cos{\frac{\theta}{2}}|0\rangle+e^{it}\sin{\frac{\theta}{2}}|1\rangle.
\end{eqnarray}

Therefore, an arbitrary unit product vector $|\mu_A\rangle\nu_B\rangle$ for two qubits can be written as
\begin{eqnarray}
&&|\mu_A\rangle\nu_B\rangle\nonumber\\
&=&(\cos{\frac{\theta_1}{2}}|0\rangle+e^{it_1}\sin{\frac{\theta_1}{2}}|1\rangle)(\cos{\frac{\theta_2}{2}}|0\rangle+e^{it_2}\sin{\frac{\theta_2}{2}}|1\rangle)\nonumber\\
&=&\cos{\frac{\theta_1}{2}}\cos{\frac{\theta_2}{2}}|00\rangle+e^{it_2}\cos{\frac{\theta_1}{2}}\sin{\frac{\theta_2}{2}}|01\rangle+\nonumber\\
&&e^{it_1}\sin{\frac{\theta_1}{2}}\cos{\frac{\theta_2}{2}}|10\rangle+e^{i(t_1+t_2)}\sin{\frac{\theta_1}{2}}\sin{\frac{\theta_2}{2}}|11\rangle.
\end{eqnarray}
\begin{eqnarray}
&&\langle\mu_A\langle\nu_B|\nonumber\\
&=&(\cos{\frac{\theta_1}{2}}\langle0|+e^{-it_1}\sin{\frac{\theta_1}{2}}\langle1|)(\cos{\frac{\theta_2}{2}}\langle0|+e^{-it_2}\sin{\frac{\theta_2}{2}}\langle1|)\nonumber\\
&=&\cos{\frac{\theta_1}{2}}\cos{\frac{\theta_2}{2}}\langle00|+e^{-it_2}\cos{\frac{\theta_1}{2}}\sin{\frac{\theta_2}{2}}\langle01|+\nonumber\\
&&e^{-it_1}\sin{\frac{\theta_1}{2}}\cos{\frac{\theta_2}{2}}\langle10|+e^{-i(t_1+t_2)}\sin{\frac{\theta_1}{2}}\sin{\frac{\theta_2}{2}}\langle11|.
\end{eqnarray}
By Eq. (\ref{exampleKorbicz}),
\begin{eqnarray}
 W&=&a|00\rangle\langle00|+a|11\rangle\langle11|+b|01\rangle\langle01|+b|10\rangle\langle10|\nonumber\\
 &&+(a+b)|01\rangle\langle10|+(a+b)|10\rangle\langle01|
\end{eqnarray}
and
\begin{eqnarray}
&&\text{tr}[W(|\mu_A\rangle\nu_B\rangle\langle\mu_A\langle\nu_B|)]\\
&=&a\cos^2{\frac{\theta_1}{2}}\cos^2{\frac{\theta_2}{2}}+b\cos^2{\frac{\theta_1}{2}}\sin^2{\frac{\theta_2}{2}}\nonumber\\
&&+(a+b)\cos{\frac{\theta_1}{2}}\sin{\frac{\theta_2}{2}}\sin{\frac{\theta_1}{2}}\cos{\frac{\theta_2}{2}}e^{i(t_1-t_2)}\nonumber\\
&&+(a+b)\sin{\frac{\theta_1}{2}}\cos{\frac{\theta_2}{2}}\cos{\frac{\theta_1}{2}}\sin{\frac{\theta_2}{2}}e^{i(t_2-t_1)}\nonumber\\
&&+b\sin^2{\frac{\theta_1}{2}}\cos^2{\frac{\theta_2}{2}}+a\sin^2{\frac{\theta_1}{2}}\sin^2{\frac{\theta_2}{2}}\\
&=&\frac{1}{2}(a+b)(\cos^2{\frac{\theta_1}{2}}\cos^2{\frac{\theta_2}{2}}+\cos^2{\frac{\theta_1}{2}}\sin^2{\frac{\theta_2}{2}}\nonumber\\
&&+\sin^2{\frac{\theta_1}{2}}\cos^2{\frac{\theta_2}{2}}+\sin^2{\frac{\theta_1}{2}}\sin^2{\frac{\theta_2}{2}})\nonumber\\
&&+(a+b)\cos{\frac{\theta_1}{2}}\sin{\frac{\theta_2}{2}}\sin{\frac{\theta_1}{2}}\cos{\frac{\theta_2}{2}}e^{i(t_1-t_2)}\nonumber\\
&&+(a+b)\sin{\frac{\theta_1}{2}}\cos{\frac{\theta_2}{2}}\cos{\frac{\theta_1}{2}}\sin{\frac{\theta_2}{2}}e^{i(t_2-t_1)}\nonumber\\
&&+\frac{1}{2}(a-b)(\cos^2{\frac{\theta_1}{2}}\cos^2{\frac{\theta_2}{2}}-\cos^2{\frac{\theta_1}{2}}\sin^2{\frac{\theta_2}{2}}\nonumber\\
&&-\sin^2{\frac{\theta_1}{2}}\cos^2{\frac{\theta_2}{2}}+\sin^2{\frac{\theta_1}{2}}\sin^2{\frac{\theta_2}{2}})\\
&=&\frac{1}{2}(a+b)(\cos^2{\frac{\theta_1}{2}}\cos^2{\frac{\theta_2}{2}}+\cos^2{\frac{\theta_1}{2}}\sin^2{\frac{\theta_2}{2}}\nonumber\\
&&+\sin^2{\frac{\theta_1}{2}}\cos^2{\frac{\theta_2}{2}}+\sin^2{\frac{\theta_1}{2}}\sin^2{\frac{\theta_2}{2}})\nonumber\\
&&+(a+b)\cos{\frac{\theta_1}{2}}\sin{\frac{\theta_2}{2}}\sin{\frac{\theta_1}{2}}\cos{\frac{\theta_2}{2}}e^{i(t_1-t_2)}\nonumber\\
&&+(a+b)\sin{\frac{\theta_1}{2}}\cos{\frac{\theta_2}{2}}\cos{\frac{\theta_1}{2}}\sin{\frac{\theta_2}{2}}e^{i(t_2-t_1)}\nonumber\\
&&+\frac{1}{2}(a-b)\cos{\theta_1}\cos{\theta_2}\label{NonEW}.
\end{eqnarray}

Let $t_1-t_2=\pi$ in Eq. (\ref{NonEW}). We have
\begin{eqnarray}
&&\text{tr}[W(|\mu_A\rangle\nu_B\rangle\langle\mu_A\langle\nu_B|)]\nonumber\\
&=& \frac{1}{2}(a+b)(\cos{\frac{\theta_1}{2}}\cos{\frac{\theta_2}{2}}-\sin{\frac{\theta_1}{2}}\sin{\frac{\theta_2}{2}})^2\nonumber\\
&&+\frac{1}{2}(a+b)(\cos{\frac{\theta_1}{2}}\sin{\frac{\theta_2}{2}}-\sin{\frac{\theta_1}{2}}\cos{\frac{\theta_2}{2}})^2\nonumber\\
&&+\frac{1}{2}(a-b)\cos{\theta_1}\cos{\theta_2}\label{NonEW1}.
\end{eqnarray}

Let $\theta_1=\theta_2=\pi/2$ in Eq. (\ref{NonEW1}). We have
\begin{eqnarray}
&&\text{tr}[W(|\mu_A\rangle\nu_B\rangle\langle\mu_A\langle\nu_B|)]=0.
\end{eqnarray}

Let $t_1=\pi, t_2=0, \theta_1=\theta_2=\pi/2$. We have $|\mu_A\rangle=\frac{\sqrt2}{2}(|0\rangle-|1\rangle), |\nu_B\rangle=\frac{\sqrt2}{2}(|0\rangle+|1\rangle)$.

\section*{Appendix B: The procedure of the proof of the optimality of an EW in the two-qubit case}
\textbf{Proof:}
Suppose an arbitrary qubit entangled state
\begin{eqnarray}
\rho=\left[\begin{array}{cccc} \rho_{00} & \rho_{01} & \rho_{02} & \rho_{03}\\
                               \rho_{10} & \rho_{11} & \rho_{12} & \rho_{13}\\
                               \rho_{20} & \rho_{21} &\rho_{22} & \rho_{23}\\
                               \rho_{30} & \rho_{31} & \rho_{32} & \rho_{33} \end{array}\right].
\end{eqnarray}

By Eq. (\ref{exampleKorbicz}), we have
\begin{eqnarray}
&&\text{tr}(W\rho)\\
&=&a\rho_{00}+b\rho_{11}+(a+b)(\rho_{12}+\rho_{21})+b\rho_{22}+a\rho_{33}\\
&=&a(\rho_{00}+\rho_{12}+\rho_{21}+\rho_{33})+b(\rho_{11}+\rho_{12}+\rho_{21}+\rho_{22}).
\end{eqnarray}

Since $\rho$ is positive, $\rho_{11}+\rho_{12}+\rho_{21}+\rho_{22}\geq0$. If $\text{tr}(Q_2^\Gamma\rho)=a\rho_{00}+a\rho_{12}+a\rho_{21}+a\rho_{33}<0$, $\text{tr}(W\rho)<0$. The EW $Q_2^\Gamma$ is finer than the EW $W$ either for $b>a$ or for $b\leq a$.

On the other hand, (unnormalized) $Q_2^\Gamma$ is the same EW as $W=|\psi\rangle\langle\psi|^\Gamma$, the optimal EW whose SPA is separable \cite{Augusiak11}, where $\psi=\frac{1}{\sqrt2}(|00\rangle+|11\rangle)$.\hfill$\blacksquare$


\begin{thebibliography} {99}
\bibitem{Thanks} We are indebted to the anonymous for bringing this point to our attention.
\end{thebibliography}

\end{document}